\documentclass[3p,times]{elsarticle}

\newcommand{ \be }{\begin{equation}}
\newcommand{ \ee }{\end{equation}}
\newcommand{ \bea }{\begin{eqnarray}}
\newcommand{ \eea }{\end{eqnarray}}
\newcommand{ \bff }{\begin{figure}[htpb]}
\newcommand{ \ef }{\end{figure}}
\newcommand{ \bmn }{\begin{minipage}}
\newcommand{ \emn }{\end{minipage}}
\newcommand{ \bt }{\begin{table}[htpb]}
\newcommand{ \et }{\end{table}}

\newcommand{ \pt }{$p_{T}$}

\newcommand{\epem}{$e^{+}e^{-}$}

\usepackage{ecrc}

\usepackage{epstopdf}

\volume{00}

\firstpage{1}

\journalname{Nuclear Physics A}

\runauth{C. Yang et al.}

\jid{nupha}

\jnltitlelogo{Nuclear Physics A}

\usepackage{graphicx}
\usepackage{amsmath,amssymb}

\usepackage{lineno}

\begin{document}

\begin{frontmatter}



\title{Direct photon production in Au+Au collisions at $\sqrt{s_{NN}}=200$ GeV at STAR}

\author{Chi Yang (for the STAR\fnref{col1} Collaboration)}
\fntext[col1] {A list of members of the STAR Collaboration and acknowledgements can be found at the end of this issue.}
\address{Department of Modern Physics, University of Science and Technology of China, Hefei, Anhui 230026, China}
\address{Brookhaven National Laboratory, Upton, NY 11973, USA}




\begin{abstract}
We present the direct photon production for $1<p_{T}<10$ GeV/$c$ derived from continuum in the dielectron invariant mass region $0.1<M_{ee}<0.3$ GeV/$c^{2}$ from one billion $\sqrt{s_{NN}}=200$ GeV Au+Au events taken in year 2010 and 2011. A clear excess in the invariant yield compared to the number of binary collision scaled $p+p$ reference is observed in the \pt~range 1-4 GeV/$c$. Model calculations with contributions from thermal radiation and initial hard parton scattering are consistent within uncertainties with the direct photon invariant yield.
\end{abstract}


\end{frontmatter}



\section{Introduction}
\label{intro}
Direct photons and dileptons are clean probes to study the fundamental properties of a hot and dense medium created in ultra-relativistic heavy-ion collisions. Due to minimal interactions with the strongly interacting medium, they can bring the information of the whole time evolution and dynamics of the medium. The PHENIX direct-photon measurements in Au+Au, $d$+Au and $p+p$ collisions at $\sqrt{s_{NN}}=200$ GeV showed that for $p_{T}<4$ GeV/$c$ an enhanced yield above the number of binary collision ($N_{coll}$) scaled $p+p$ production is observed in Au+Au while there is no similar enhancement in $d$+Au collisions. In the high \pt~range, the invariant yield in Au+Au collisions is consistent with $N_{coll}$ scaled $p+p$ production and the NLO pQCD predictions~\cite{DVP_PHENIX_dAu,DVP_PHENIX_temp}. The results from ALICE in Pb+Pb and $p+p$ collisions at $\sqrt{s_{NN}}=2.76$ TeV showed an excess in Pb+Pb over the scaled $p+p$ for $p_{T}<4$ GeV/$c$~\cite{DVP_ALICE} while the yield measured by CMS collaboration in Pb+Pb collisions follows the scaled $p+p$ results and NLO pQCD predictions in the photon transverse energy ($E_{T}$) range 20-80 GeV ~\cite{DVP_CMS}. These measurements indicate that in the low \pt~range, there are contributions from other sources rather than initial hard parton scattering. The study on direct photons opens two kinematic windows. One is the low \pt~range for the in-medium effect study. The other one is the high \pt~range for the hard process research. With the comparison to model prediction, some key parameters of the QGP matter such as initial temperature can be obtained~\cite{private}.

There are two methods for direct photon measurement. One is the real photon method in which one measures all inclusive photons and then subtracts the hadron decayed photons from them. The other one is the virtual photon method in which one measures the virtual photon by their associate dielectron pairs and then deduces the direct photon from the relationship between virtual photon and direct photon. The latter is used in this article. The dielectron measurement in Au+Au collisions at 200 GeV at STAR from year 2010 data has been published recently~\cite{PRL}. The preliminary results on dielectron production in $p+p$ collisions at 200 GeV from year 2012 can be found in ~\cite{run12pp}. In this article, we focus on the measurement of direct photon in Au+Au
collisions at 200 GeV.

\section{Analysis}
\label{analysis}
\begin{figure}
\begin{center}
\includegraphics*[width=6.cm]{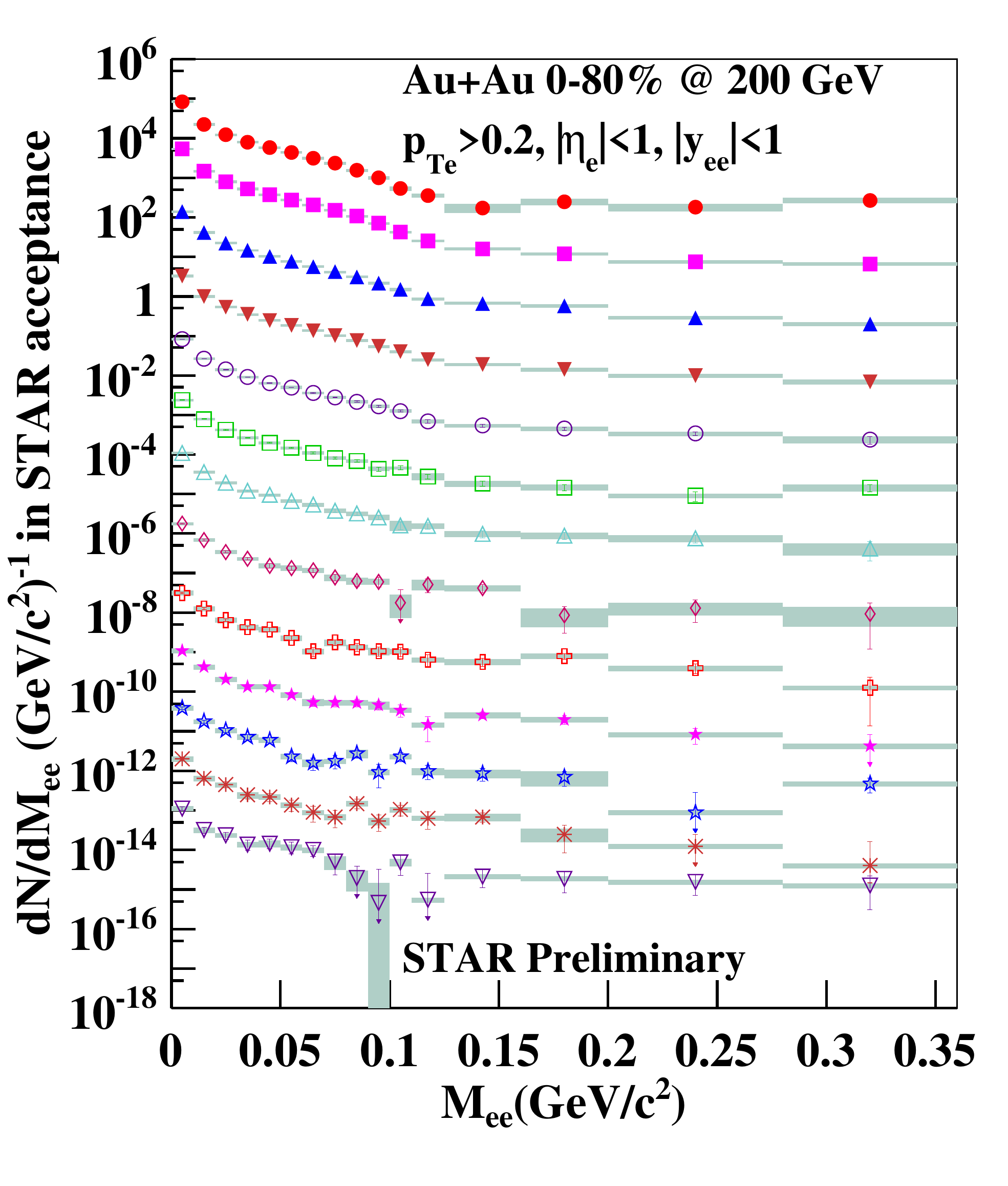}
\includegraphics*[width=3.4cm]{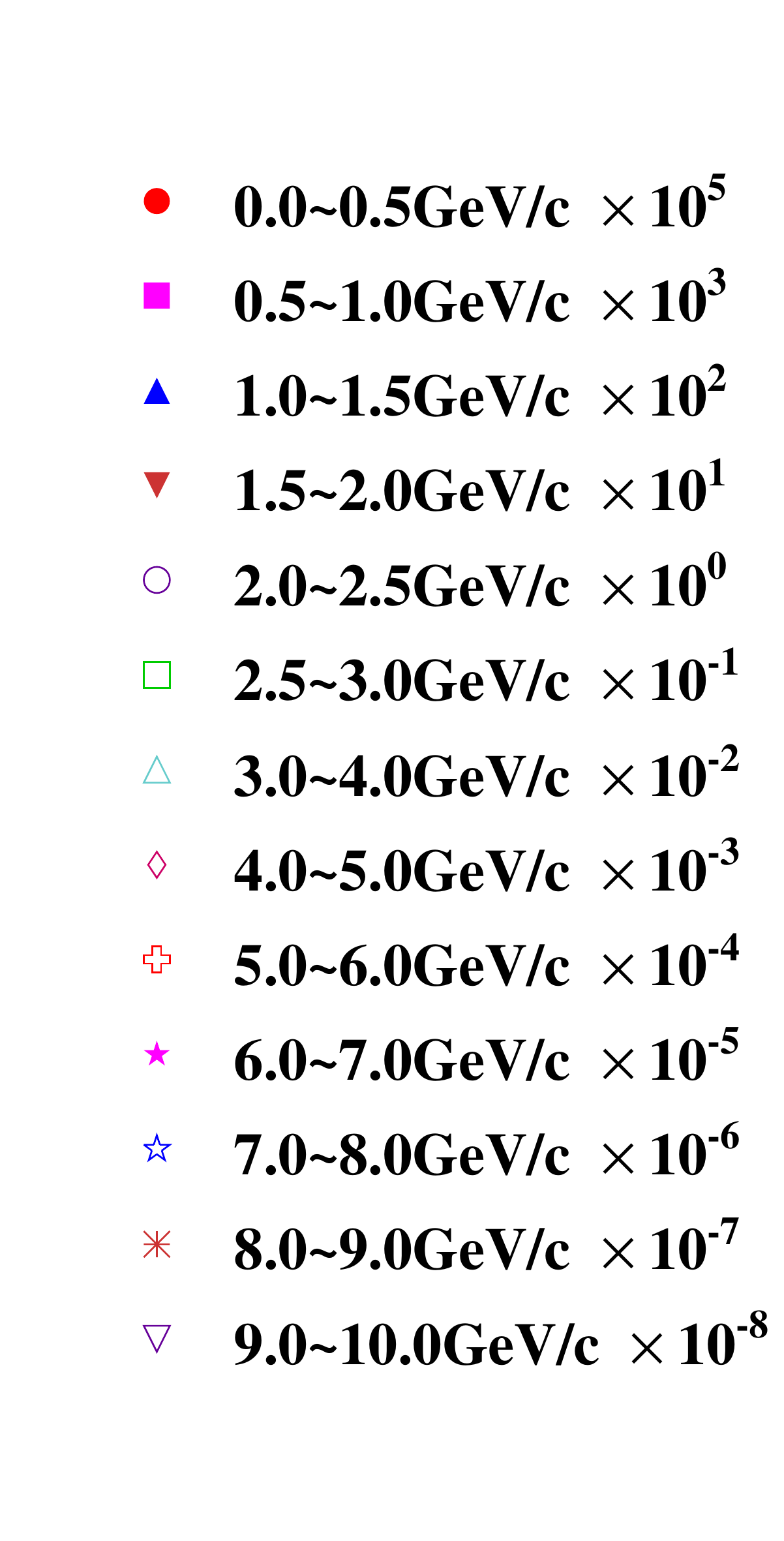}
\caption{Dielectron invariant mass spectra in the low mass range for different $p_{T}$ slices. The spectra in different \pt~slices are scaled by different factors for clarification. The error bars and the shaded bands represent the statistical and systematic uncertainties, respectively.}\label{fig:continuum}
\end{center}
\end{figure}

The data used in this analysis are from Au+Au collisions at $\sqrt{s_{NN}}=200$ GeV collected in year 2010 (Run10) and 2011 (Run11). There are 258 million and 488 million minimum bias (0-80\%) events from Run10 and Run11, respectively, passing bad run rejections and vertex selections. The main subsystems used for electron identification are the Time Projection Chamber (TPC)~\cite{TPC} and the Time-of-Flight detector (TOF)~\cite{TOF} for the minimum-bias events. With the requirements on the particle energy loss (dE/dx) measured by TPC close to electron expected dE/dx and time-of-flight measured by TOF close to the speed of light, electron samples can be selected with high purity. The electron purity is about 95\% in Au+Au minimum bias collisions. The dielectron invariant mass spectra are obtained separately with Run10 and Run11 datasets. The final results are then combined point-to-point according to their relative statistical uncertainties. To improve the statistics at high \pt, we use the Barrel Electro-Magnetic Calorimeter (BEMC)~\cite{BEMC} triggered events taken in Run11. The 39 million BEMC triggered events used in this analysis are equivalent to 5389 million minimum bias triggered events. The BEMC trigger significantly enhances the capability of STAR for high \pt~dielectron measurement. For BEMC triggered electrons, additional requirements on the ratio of momentum measured by the TPC to the energy deposited in the BEMC are utilized. The electron purity can reach about 80\% for BEMC triggered electron sample. The dielectron invariant mass spectrum in this paper are within STAR acceptance ($p_{T}^{e}>0.2$ GeV/$c$,$~|\eta^{e}|<1,~|y^{ee}|<1$) and corrected for efficiency, in which $p_{T}^{e}$ is the electron \pt, $\eta^{e}$ is the electron pseudo-rapidity, and $y^{ee}$ is the rapidity of electron-positron pairs. The dielectron invariant mass spectra in different \pt~ranges are shown in Fig.~\ref{fig:continuum}. The results for $p_{T}<5$ GeV/$c$ are the combined results from Run10 and Run11 minimum bias data. The results for $p_{T}>5$ GeV/$c$ are from the BEMC triggered data. More dielectron analysis details and results can be found in~\cite{PRL,PRC}.

A contribution from internal conversion of direct photons is expected. The relation between real photon and the associated \epem~pair production can be described as Eq.~\ref{eq:dvp2ee}~\cite{DVP2ee,Landsberg}.
\begin{equation}\label{eq:dvp2ee}
\frac{dN_{ee}}{dM} = \frac{2\alpha}{3\pi}\sqrt{1-\frac{4m_{e}^{2}}{M^{2}}}(1+\frac{2m_{e}^{2}}{M^{2}})\frac{S(M)}{M}dN_{\gamma}
\end{equation}

In Eq.~\ref{eq:dvp2ee}, M is the mass of the virtual photon or the \epem~pair, $\alpha$ is the fine structure constant, and S(M) describes the difference between real photon process and virtual photon process. In this analysis, this factor S(M) is assumed to be 1 in the range $M_{ee}<0.3$ GeV/$c^{2}$, $p_{T}>1$ GeV/$c$. Based on Eq.~\ref{eq:dvp2ee}, the measurement of dielectron spectrum in the low mass region will deduce the production of direct photons~\cite{DVP_PHENIX_detailed}.

\begin{figure}
\begin{center}
\includegraphics*[width=0.47\textwidth]{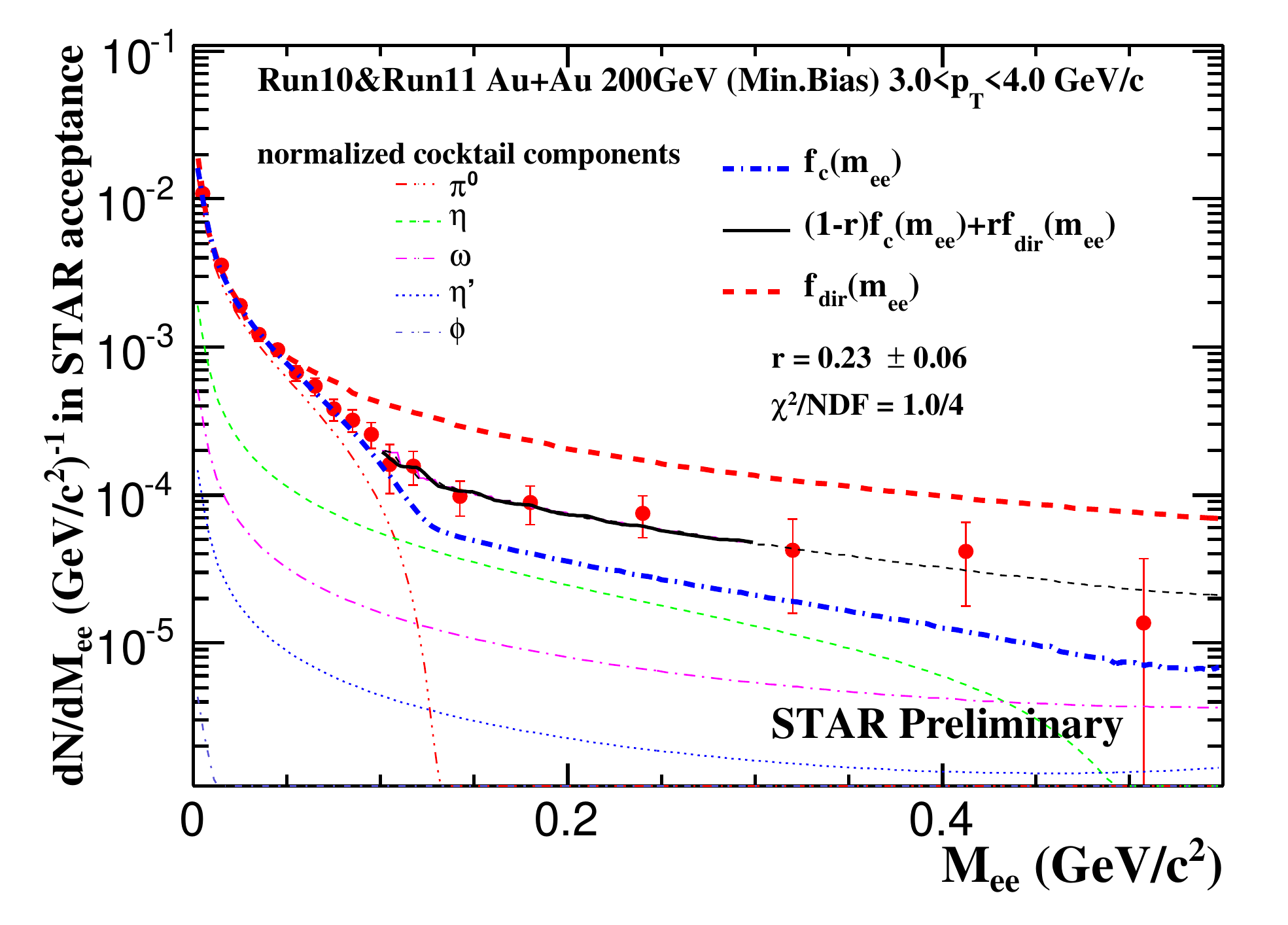}
\includegraphics*[width=0.47\textwidth]{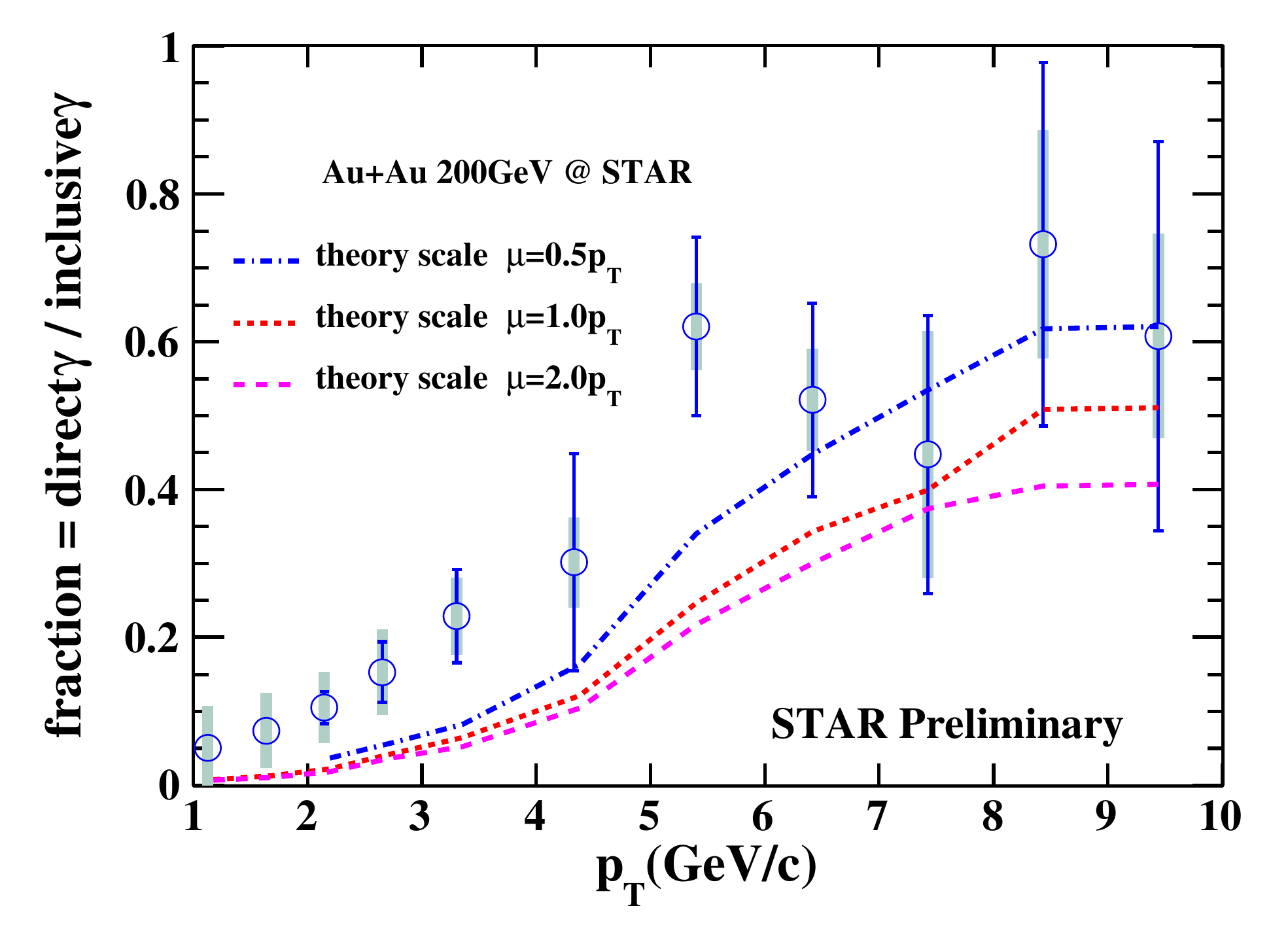}
\caption{Left panel shows the two-component function fit on Au+Au dielectron spectra for $0.1<M_{ee}<0.3$ GeV/$c^{2}$ at $3<p_{T}<4$ GeV/$c$. Red dashed line is $f_{dir}$ while blue dashed-dot line is $f_{cocktail}$. The black solid line is the fit to the data for $0.1<M_{ee}<0.3$ GeV/$c^{2}$. The black dashed line is the extrapolation of the fit function outside the fit range. Right panel shows the ratio of direct photon to the inclusive photon compared with the NLO pQCD prediction (see text). The error bars and the shaded bands represent the statistical and systematic uncertainties, respectively.}\label{fig:fitting}
\end{center}
\end{figure}

The direct photon yields are extracted by fitting the dielectron invariant mass spectra in the low mass region with two components. One component is the contribution from hadron cocktail and the other one comes from direct photon internal conversion. In the fit function $(1-r)f_{cocktail}+rf_{dir}$, $r$ is the ratio of direct photon to inclusive photon. The $f_{cocktail}$ is the normalized cocktail while the $f_{dir}$ is the normalized internal conversion from direct photon passing STAR acceptance in different $p_{T}$ ranges. The left panel of Fig.~\ref{fig:fitting} shows an example of the two-component fit. The uncertainties shown in the figure are the quadrature sum of statistical and systematic uncertainties. The blue dashed-dot line and red-dashed lines represent the normalized cocktail and internal conversion from direct photon, respectively. The black solid line is the fit to the data at $0.1<M_{ee}<0.3$ GeV/$c^{2}$. The black dashed line is the extrapolation of the fit function outside the fit range. Right panel of Fig.~\ref{fig:fitting} shows the ratio of direct photon to inclusive photon compared with the NLO pQCD prediction. The curves represent $T_{AA}d\sigma_{\gamma}^{NLO}(p_{T})/N_{\gamma}^{inclusive}(p_{T})$, showing the scale dependence of the theory~\cite{NLOpQCDcurve}, in which $T_{AA}$ is the nuclear overlapping factor. The $d\sigma_{\gamma}^{NLO}(p_{T})$ is the \pt-differential invariant cross section of direct photon obtained from~\cite{DVP_PHENIX_pp}. The $N_{\gamma}^{inclusive}(p_{T})$ is the STAR measurement which is in form of $F/(\frac{2\alpha}{3\pi})/(2\pi p_{T}dp_{T}dy)$, in which $F$ is the normalization factor of $f_{dir}$. The data show consistency with NLO pQCD calculation within uncertainties at $p_{T}>4$ GeV/$c$. A clear enhancement in data compared to the calculation for $1<p_{T}<4$ GeV/$c$ is observed.

\begin{figure}
\begin{center}
\includegraphics*[width=0.47\textwidth]{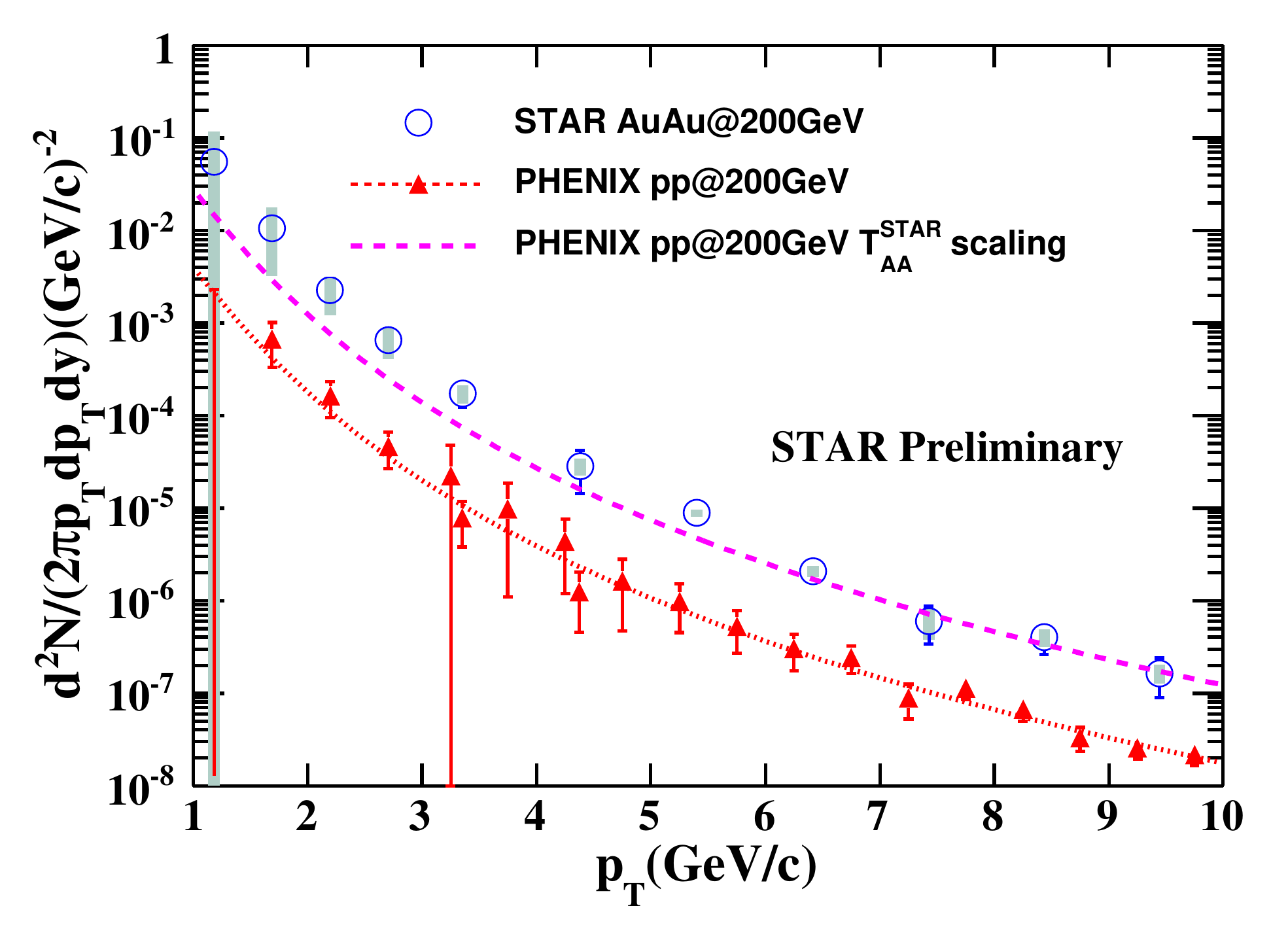}
\includegraphics*[width=0.47\textwidth]{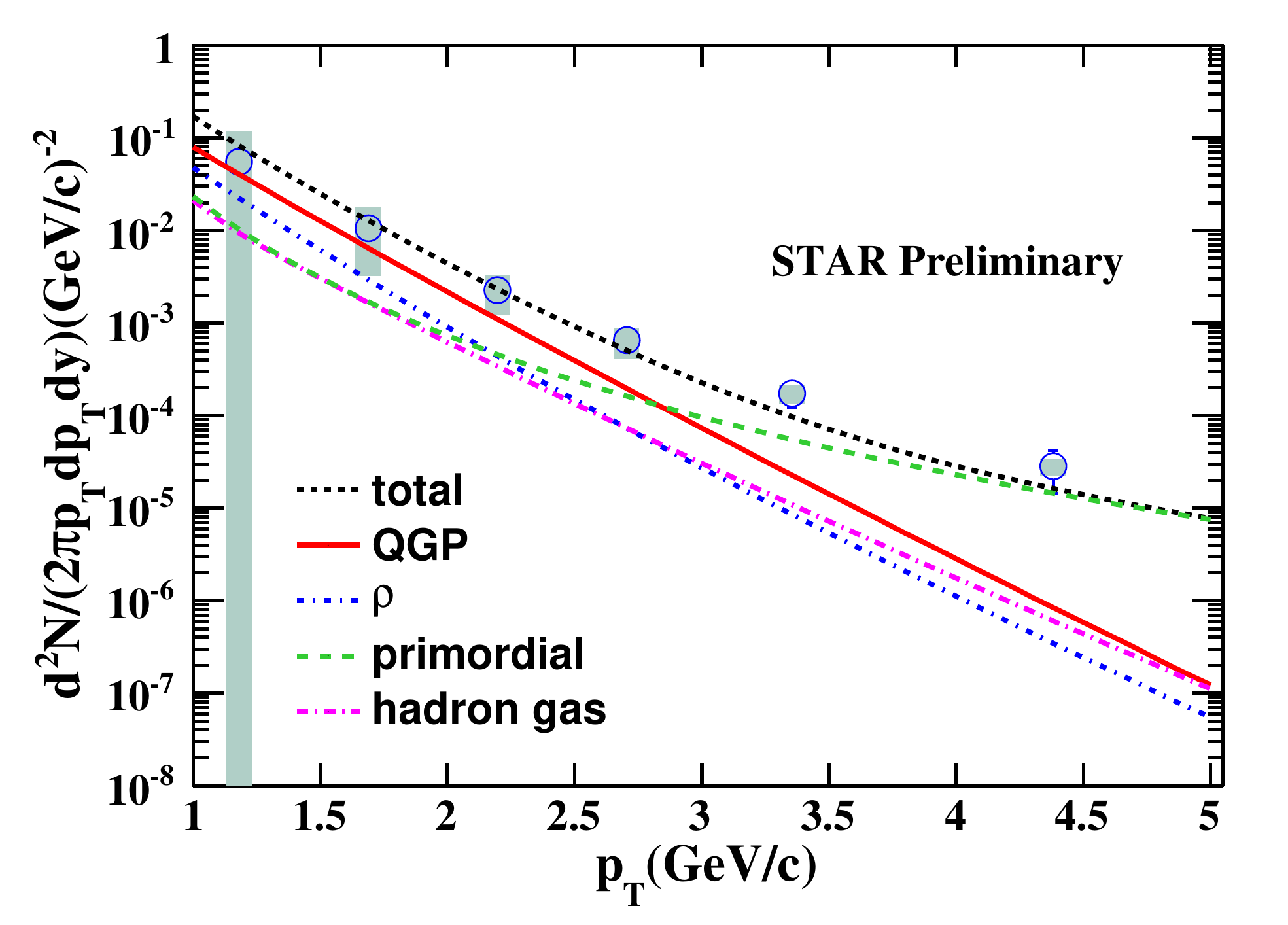}
\caption{Left panel shows the direct photon invariant yield. The pink dashed curve represents a power-law fit to PHENIX 200 GeV $p+p$ cross section~\cite{DVP_PHENIX_detailed,DVP_PHENIX_pp}, scaled by $T_{AA}$. Right panel shows the direct photon invariant yield compared to model predictions including the contributions from different sources~\cite{private,private1}. The statistical and systematic uncertainties are shown by the bars and bands, respectively.}\label{fig:yieldcompare}
\end{center}
\end{figure}

Figure~\ref{fig:yieldcompare} shows the invariant yield of the direct photons in Au+Au collisions at 200 GeV. In the left panel, the open circles are the direct photon invariant yields measured from STAR 200 GeV 0-80\% Au+Au collisions. The uncertainty is dominated by systematic uncertainty in the cocktail. The large systematic uncertainties in 1-2 GeV/$c$ are due to the lack of $\eta$ measurement in the corresponding \pt~range. The $\eta$ dN/dy and \pt~input for cocktail simulation in this range are based on the extrapolation of the TBW model~\cite{TBW}, while at $p_{T}>2$ GeV/$c$, it is constrained by the measurements. The triangles are the invariant cross sections in $p+p$ collisions at 200 GeV measured by PHENIX collaboration. They are parameterized by a power-law function, as shown in dotted line. The parameterized distribution is then scaled by the nuclear overlapping factor $T_{AA}=N_{bin}/42$mb = 6.95mb$^{-1}$ as shown in dashed line, and compared to the Au+Au results. For $1<p_{T}<4$ GeV/$c$, the Au+Au results are clearly higher than the $T_{AA}$ scaled $p+p$ results, while at high \pt~the Au+Au yield is consistent with the scaled $p+p$ expectation. The right panel shows a comparison between STAR Au+Au data and model calculations done by Ralf Rapp et al.~\cite{private,private1}. The curves ``QGP'' stands for the contribution from QGP thermal radiation. The ``$\rho$'' and ``hadron gas'' represent contributions from in-medium $\rho$ and other mesonic interactions in the hadronic gas, respectively. The ``primordial'' line depicts the contribution from the initial hard parton scattering. The comparison shows consistency between the model calculation and the measurement within uncertainties. In the low \pt~range 1-4 GeV/$c$, the dominant sources are from thermal radiation while as the \pt~increases to 4-5 GeV/$c$ the initial hard parton scattering becomes the dominant source.

\section{Conclusions}
\label{conclusion}
We measured direct photon production in Au+Au collisions at STAR at $\sqrt{s_{NN}}=200$ GeV. The direct photon measurement based on virtual photon method is firstly extended to high \pt~of 10 GeV/$c$. In the low \pt~range 1-4 GeV/$c$, the production of direct photon can access the thermal radiation from QGP and hadron gas. The direct photon invariant yield in this range shows a clear excess in Au+Au over the number of binary collision scaled $p+p$ results. In the \pt~range above 4 GeV/c, there is no clear enhancement observed. A model prediction which includes the contributions from thermal radiation and initial hard process is consistent with our direct photon yield within uncertainties.









\end{document}